\documentclass[12pt]{iopart}

\usepackage{graphics}
\usepackage{graphicx}
\usepackage{subfig}
\usepackage{amssymb}
\usepackage{float}
\usepackage{color}
\begin{document}

\title[Bottle-brush brushes under good solvent conditions]{Structure of bottle-brush 
brushes under good solvent conditions. A molecular dynamics study}

\author{Hamed Maleki$^1$ and Panagiotis E Theodorakis$^2$$^,$$^3$$^,$$^4$}
                                                                                                 
\address{$^1$ Institut f\"ur Physik, Johannes Gutenberg-Universit\"at, Staudinger Weg 7, D-55099 Mainz, Germany}
\address{$^2$ Faculty of Physics, University of Vienna, Boltzmanngasse 5, A-1090 Vienna, Austria}
\address{$^3$ Institute for Theoretical Physics and Center for Computational Materials Science, Vienna University of Technology, Hauptstra{\ss}e 8-10, A-1040 Vienna, Austria}
\address{$^4$ Vienna Computational Materials Laboratory, Sensengasse 8/12, A-1090 Vienna, Austria}

\ead{maleki@uni-mainz.de}
\ead{panagiotis.theodorakis@univie.ac.at}
                                                                                                 
\date{\today}    

%
\begin{abstract}                                                                                 
We report a simulation study for bottle-brush polymers grafted on a 
rigid backbone. Using a standard
coarse-grained bead-spring model extensive molecular dynamics simulations for
such macromolecules under good solvent conditions are
performed. We consider a broad range of parameters and present 
numerical results for the monomer density profile, density of the untethered
ends of the grafted flexible backbones and 
the correlation function describing the range that neighboring grafted 
bottle-brushes are affected by the presence of the others due to the 
excluded volume interactions. The end beads of the 
flexible backbones of the grafted bottle-brushes
do not access the region close to the rigid 
backbone due to the presence of the side chains of the grafted 
bottle-brush polymers, which stretch further the chains in the radial
directions. 
Although a number of different correlation lengths exist as a result of  the complex structure of these macromolecules,
their properties can be tuned with high accuracy in good solvents.
Moreover, qualitative differences with "typical" bottle-brushes are discussed.
Our results provide a first approach to characterizing such complex macromolecules
with a standard bead spring model.
\end{abstract} 

\submitto{\JPCM}
%
\maketitle
%
\section{Introduction}

Macromolecules with comb-like architecture, where linear or branched 
side chains are grafted onto a backbone chain 
have found much interest in recent years due to their physical and 
biochemical properties, which offer important benefits
~\cite{zhang_cylindrical_2005,koutalas_well-defined_2005,
subbotin_spatial_2007,ishizu_architecture_2008,sheiko_cylindrical_2008,
potemkin_comblike_2009, schappacher_newpolymerchain_2000}. 
Although their structure is complicated, such macromolecules are being synthesized 
rather successfully~\cite{zhang_cylindrical_2005,koutalas_well-defined_2005, 
ishizu_architecture_2008,wintermantel_molecular_1996,
gunari_surfactant-induced_2008}.
Atom transfer radical polymerization has been 
efficiently applied to the synthesis of various bottle-brush 
brushes (also known as comb-on-comb brushes)~\cite{koutalas_well-defined_2005,
ishizu_architecture_2008,sheiko_cylindrical_2008,matyjaszewski_atom_2001, 
matyjaszewski_effect_2003}, where bottle-brush 
molecules are grafted onto a linear chain (or a point (single monomer) in the case of 
star bottle-brush brushes) which serves as the backbone of the macromolecule. 
Brush polymers have been studied extensively as far as it concerns experiments
~\cite{zhang_cylindrical_2005,koutalas_well-defined_2005,
subbotin_spatial_2007,ishizu_architecture_2008,sheiko_cylindrical_2008,
potemkin_comblike_2009,11,12,13} as well as theoretical modeling~\cite{alexander_adsorption_1977,
de_gennes_conformations_1980,daoud_star_1982,bug_theory_1987,
milner_parabolic_1988,milner_effects_1989,milner_polymer_1991,
fredrickson_surfactant-induced_1993,zhulina_scaling_1995}. 
Already, the study of ''typical'' bottle-brush 
polymers (flexible linear polymeric chains grafted onto a backbone) 
has attracted attention for potential
applications~\cite{zhang_cylindrical_2005,sheiko_cylindrical_2008}.
For example, some work was motivated by the use of these cylindrical 
brushes as building blocks in functional supramolecular structures; 
applications for actuators and sensors have been also discussed, since the 
structure of these stimuli-responsive polymers can change when 
external parameters such as pH of the solution, temperature etc. are 
varied~\cite{stephan_shape_2002,li_new_2004}. The study of such 
effects has already been the subject of previous simulation studies of 
bottle-brush macromolecules with one or two types of side 
chains where a rigid backbone has been considered as a starting
point for the study of these complex macromolecules~\cite{theodorakis_microphase_2009,theodorakis_interplay_2010,
theodorakis_pearl-necklace_2010, theodorakis_jcp_2011, theodorakis_epje_2011, hsu_theodorakis_jcp_2011, hsu_characteristic_2010, 
hsu_one-_2007,hsu_structure_2008,hsu_intramolecular_2006, hsu_how_2009}, where
it has also been shown that the radial density profiles for bottle-brush 
polymers with stiff and flexible backbones are similar when good solvent
conditions are assumed.                                                                
Bottle-brush brushes could serve as a candidate for applications 
already suggested for bottle-brush polymers, but in the case of 
bottle-brush brush polymers the higher number of varied structural 
parameters could help tuning the resulting properties of such 
macromolecules with a higher accuracy and in various ways. Also, towards 
understanding more complex structures of macromolecules which exist in             
biological systems, simulation models would allow for a better 
understanding of their complex structure-properties 
relation~\cite{klein_chemistry:_2009,chang_structural_2009}.

Bottle-brush brushes exhibit rich structural behaviour 
providing even ideas for new applications due to the 
higher multitude of the correlation lengths which is expected from 
their structure. More interesting formations could be expected under
poor solvent conditions as for typical bottle-brushes  ~\cite{theodorakis_microphase_2009,theodorakis_interplay_2010,
theodorakis_pearl-necklace_2010, theodorakis_jcp_2011, theodorakis_epje_2011},
but for bottle-brush brushes more difficulties, such as equilibrating such
a systems, could be expected.
Moreover, the interlay of the local conformation of 
side chains and the global configuration of the backbone adds more 
complexity to their intricate behaviour. The interpretation of 
experimental data on bottle-brush brushes is expected to be 
controversial as is already shown in the case of bottle-brushes with 
flexible linear side chains ~\cite{hsu_characteristic_2010,hsu_one-_2007}. Also, simulations has shown that the 
application of scaling concepts in the latter case is proven 
problematic even in the case where only a single type of side chains 
occurs, and good solvent conditions are assumed~\cite{hsu_characteristic_2010, 
hsu_one-_2007,hsu_structure_2008,hsu_intramolecular_2006,hsu_how_2009}.
Moreover, the application of theoretical arguments for the 
description of these macromolecules has been also proven
very difficult ~\cite{zhang_cylindrical_2005, subbotin_spatial_2007,
sheiko_cylindrical_2008,potemkin_comblike_2009}. 
A comparison of the results of computer simulations could rather be
realised on the basis of effective exponents, as 
for ''typical'' bottle brushes ~\cite{hsu_theodorakis_jcp_2011}. In view of the 
experimental interest for bottle-brush brushes and related
complex macromolecules some fundamental 
understanding on their static properties can be achieved rather well 
by computer simulations. In this work we describe large-scale molecular 
dynamics simulations of an off-lattice model of a bottle-brush brush 
under good solvent conditions to provide a first approach to 
understanding their overall structural properties.

The outline of this paper is as follows. In Sec. 2 we describe the 
simulation model. Then, Section 3 discusses the analyzed properties 
and our numerical results, while Section 4 presents and summarizes 
our conclusions.
%
%
\section{Computational details and simulation method}

\begin{figure*}
\begin{center}
\rotatebox{0}{\resizebox{!}{0.50\columnwidth}{%
  \includegraphics{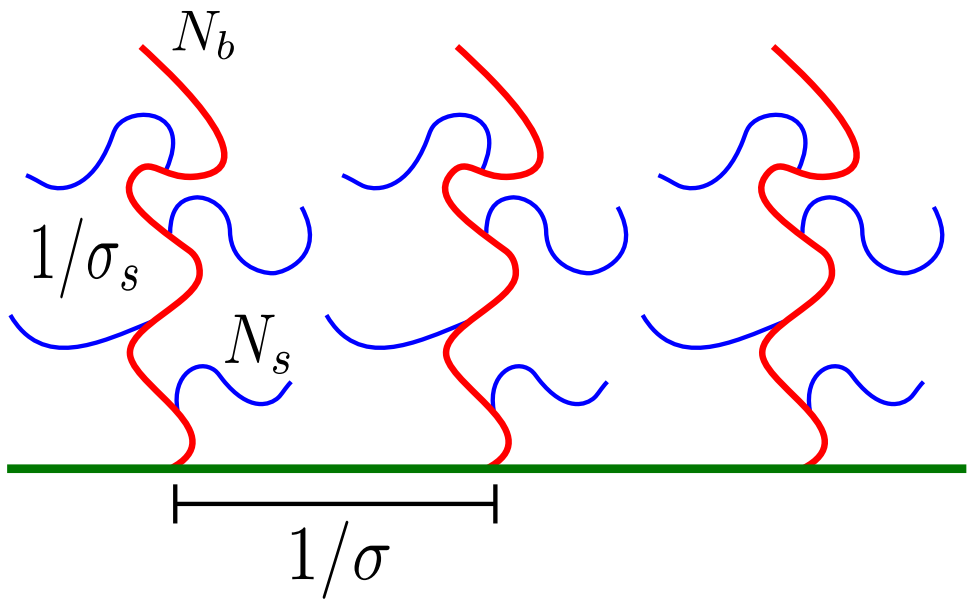}
}}
\end{center}
\caption{\label{fig1} 
(Colour online) Schematic representation of bottle-brush brush polymer and parameters
describing the structure of such system.
$\sigma$ is the grafted density that the bottle brush polymers are grafted onto
the rigid backbone, $N_{b}$ is the length of the flexible backbone. 
Also, flexible side chains of length $N_{s}$ are grafted onto the flexible backbone 
with a grafting density $\sigma_{s}$.
}
\end{figure*}

\begin{figure*}
\begin{center}  
\rotatebox{0}{\resizebox{!}{0.36\columnwidth}{%
  \includegraphics{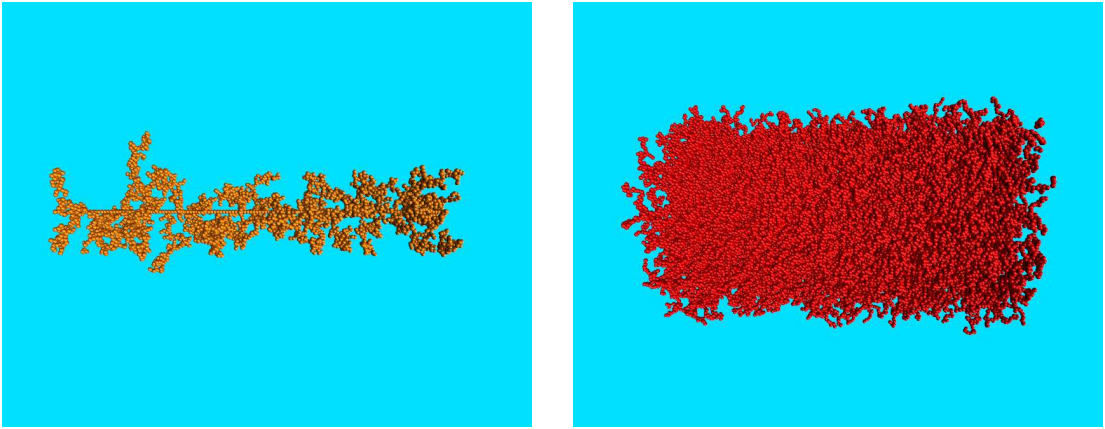}                                                                
}}                                                                                               
\end{center}                                                                                     
\caption{\label{fig2}
(Colour online) Two characteristic snapshots of bottle-brush brush polymers with 
different set of parameters. These snapshots show the different 
structures of bottle-brush brush polymers under good solvent conditions. 
This is similar to the behaviour 
of typical bottle-brush polymers. In the left snapshot the set of 
parameters $\sigma=0.25$, $N_{b}=12$, $N_{s}=3$ and $\sigma_{s}=0.50$ 
was chosen and for the right snapshot, $\sigma=1.00$, $N_{b}=48$, $N_{s}=24$, 
$\sigma_{s}=1.00$. Our choice of parameters show the different structures
that may result ranging from low densed brushes to homogeneous cylinders.
}
\end{figure*}

To simulate a bottle-brush brush polymer, we use a molecular dynamics 
method, where the monomers are coupled to a heat bath~\cite{grest_molecular_1986}. 
All the monomeric units  are modeled by the standard bead-spring 
model~\cite{graessley_excluded-volume_1999,40, grest_structure_1993,
murat_polymers_1991}, where all monomers are treated as beads of mass
$m$. This model has been extensively used in previous simulations of 
brush-like systems and its detail discussion has been given elsewhere~\cite{40}. 
Here, we give a brief description of the parameters we use in our 
simulations. For the interaction between any two monomers a 
truncated and shifted Lennard-Jones potential acts given by
\begin{equation}
\label{cases}
U_{LJ} (r)=\cases{4 \epsilon_{LJ}[(\frac{\sigma _{LJ}}{r})^{12}-(\frac{\sigma _{LJ}}{r})^6 ] +C&for $r \leq r_{c}$\\
0&for $r>r_{c}$\\}
\end{equation}

where $r_c=2^{1/6}\sigma_{LJ}$ is the cut-off of the potential, and 
the constant $C$ is defined such that $U_{LJ}(r=r_c)$ is continuous 
at the cut-off. Henceforth units are chosen such that $\epsilon_{LJ}=1,
\sigma _{LJ}=1, k_B=1$, and the mass $m$ of the beads is also taken 
as unity. The connectivity of the beads is guaranteed by the "finitely 
extensible nonlinear elastic" (FENE) potential:
\begin{equation}\label{eq2}
 U_{\rm FENE}(r)=-\frac{1}{2} k r_{0}^{2}\ln[1-(\frac{r}{r_{0}})^{2}] 
\qquad 0<r\leq r_{0},
\end{equation}                                                                                   
where $r_{0}=1.5$, $k=30$ and $U_{FENE}(r)=\infty$ outside the range 
written in equation~(\ref{eq2}). The equation of motion for each bead reads
\begin{equation}\label{eq3}
 m\frac{d^{2}\textbf{r}_{i}}{dt^{2}}=-\nabla U_{i}- 
\gamma \frac{d\textbf{r}}{dt}+\Gamma_{i} (t).
\end{equation}
In this equation $\gamma$ is the bead friction, $\Gamma_{i}(t)$ 
describes the random force of the heat bath and $U_{i}$ is the 
potential each bead experiences due to the presence of the other
beads, when they are below the cut-off distance. The random forces $\Gamma_{i}(t)$  
satisfy the standard fluctuation-dissipation relation 
\begin{equation}\label{eq4}
<\Gamma_{i}(t).\Gamma_{j}(t^{'})>=6k_{B}T\gamma\delta_{ij}\delta(t-t^{'})
\end{equation}
where $T$ is the temperature and $k_{B}$ is the Boltzmann constant. 
Following previous work \cite{grest_molecular_1986,
graessley_excluded-volume_1999,40,grest_structure_1993,murat_polymers_1991,86,
murat_structure_1989,grest_grafted_1994} $\gamma=0.5 $ and 
$T=1.2$. Here, $\tau=(m_{LJ}\sigma_{LJ}^{2}/
\epsilon_{LJ})^{1/2}$ is the natural time unit, with units that have
been given above. We use the 
the molecular dynamics package LAMMPS~\cite{lammps} to simulate our systems
where the equations of motion for each bead (equation ~\ref{eq3}) are integrated with the velocity-Verlet 
scheme~\cite{plimpton_fast_1995} with a time step $\Delta t=0.008$.
Periodic boundary conditions in the $x$-direction are
applied, which is the axis of the rigid backbone where the brush 
chains with a flexible backbone are grafted regularly with a 
distance $1/\sigma$ between them. In this study we have considered 
different grafting densities, i.e., $\sigma=0.25, 0.50$, and $1.00$, which
corresponds to grafting every bead, every second bead, and every forth
bead of the rigid backbone. Smaller densities for our range of chain lengths
would suppress any effects due to the density, while higher grafting densities
than the ones we consider here would impose difficulties in
equilibrating our systems. 
In $y-$ and $z-$directions, periodic boundary conditions are applied 
as well, but the considered linear dimensions of the  simulation box 
were chosen large enough, so that never any interaction of the grafted
bottle brush polymers with their periodic images could occur.                            
We have considered a variety of parameters  as they are schematically described in
figure~\ref{fig1}. The backbone where the chains are grafted with
a grafting density $\sigma$ is rigid, while the backbones of 
the grafted brushes onto this rigid backbone are flexible as
well as their side chains.
The number of monomers of the rigid backbone was $N=128$. We have also 
performed simulations for other choices of backbone lengths in order 
to check the influence 
of periodic boundary conditions on the resulting properties. We found that 
our results are not affected by the presence of periodic boundary conditions
for the rigid backbone length that is taken here.
However, if one tries to simulate such a system under poor solvent conditions,
then one would expect an influence of the periodic boundary conditions
on the properties of the grafted brushes ~\cite{theodorakis_microphase_2009,theodorakis_interplay_2010,
theodorakis_pearl-necklace_2010,theodorakis_jcp_2011,theodorakis_epje_2011}. In this manuscript, we 
will present only results for $N=128$.
The backbone of the bottle-brushes attached onto the rigid backbone is
$N_{b}=12,24,36$, and $48$ (the part of the chains denoted with red color in figure ~\ref{fig1}).
Higher lengths are prohibitely difficult 
to study. Also, it would be difficult for experiments to access
such lengths as it has been discussed already for typical brushes ~\cite{hsu_characteristic_2010}. 
For the side chains of the grafted bottle-brushes onto the rigid backbone
we have typically considered chain lengths shorter than $N_{b}$, i.e., 
$N_{s}=3, 6, 12$, and $24$. 
These shorter side chains (denoted with blue color in  figure~\ref{fig1}) are 
grafted onto the flexible backbone of the brushes with a grafting 
density $\sigma_{s}=0.5$ or $1.0$. As shorter chain lengths are used for $N_{s}$,
we did not study the grafting density  $\sigma_{s}=0.25$.  Figure~\ref{fig2} shows (left part) 
typical snapshots for the case of $\sigma=0.25$, $N_{b}=48$, $N_{s}=3$
and $\sigma_{s}=0.5$. In the right part 
the case of $\sigma=1.00$, $N_{b}=48$, $N_{s}=24$ and $\sigma_{s}=0.5$ is shown.
For our choice of parameters we observe structures ranging from low
densed brushes to homogeneous cylinders.

We note here that equilibration of our systems required running our simulations for every brush
for a time range of $20\times10^{6}$ MD time-steps. Then we collect $2000$ samples
running the simulations $4\times10^{6}$ MD steps further.
By studying correlations functions for the
structural properties of the brushes that vary slowly with time, we find out
that this effort was enough to obtain reliable results. 
In the following, we present our results discussing overall structural properties
for these complex macromolecules providing an insight for the overall behavior
of these polymers.

%
%
\section{Results and Discussion}                                                                 
                                                                                                 
\begin{figure*}                                                                                  
\begin{center}                                                                                   
\rotatebox{0}{\resizebox{!}{0.50\columnwidth}{%
  \includegraphics{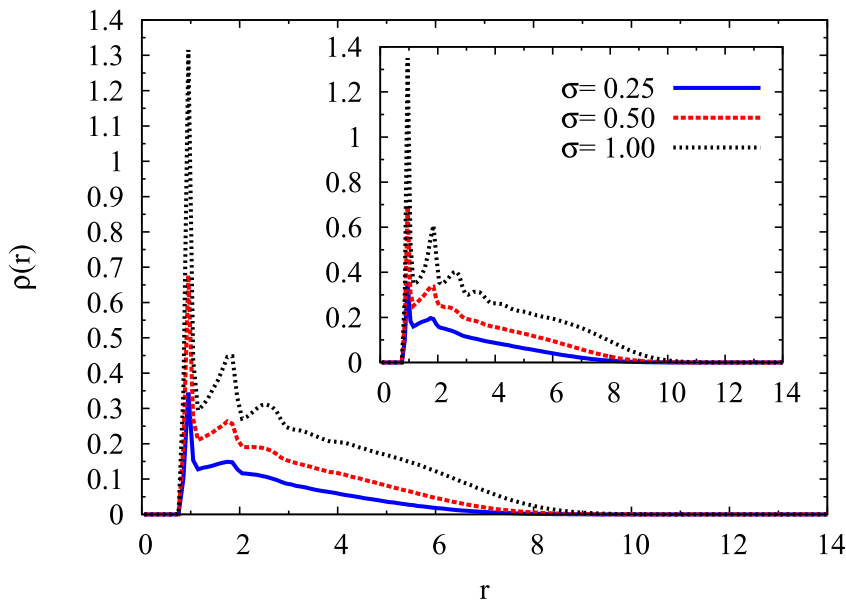}                                                                
}}                                                                                               
\end{center}                                                                                     
\caption{\label{fig3}
(Colour online) Density profile $\rho(r)$ as a function of $r$ perpendicular to the 
rigid backbone for bottle-brush brush polymers with 
$N_{b}=12$, $N_{s}=3$ and $\sigma_{s}=0.5$, for the grafting densities 
$\sigma=0.25, 0.5$ and $1.0$. $r$ is the distance from 
the rigid backbone. Inset shows results for the same set of 
parameters, but $\sigma_{s}=1.0$
}           
\end{figure*}

First, we focus on the effect of the grafting density $\sigma$ on the
density profile in the radial directions ($y-z$ plane with the rigid backbone
extending along the $x$ direction) for moderate values
of $N_{b}$ and $N_{s}$.
Figure~\ref{fig3} shows the density profiles of the system in radial 
directions ($y,z$) for the case $N_{b}=12$, $N_{s}=3$ and different 
grafted densities $\sigma$.
From figure~\ref{fig3}, the increase of the grafting density 
$\sigma$ shows that the chains overally become more stretched in the 
radial direction as it normally happens for typical bottle-brushes.
For low grafting density 
$\sigma$, the density curves decay almost exponentially for $\sigma=0.25$ and 
their height reflects analogously the corresponding grafting densities (the second 
peak from the left). Also, by increasing the grafting density $\sigma$ 
the latter extension in density becomes higher as the side chains 
of the grafted brushes get stiffened due to the excluded volume interactions. 
Moreover, for $\sigma=1.0$, figure~\ref{fig3} shows that 
the density profile persists over longer distances from the 
rigid backbone and indicates that the side chains of the 
grafted bottle-brush molecules tend to stretch in the radial 
directions. In addition, the increase of number of side chains grafted onto the 
flexible backbone $\sigma_{s}$ (inset) hint the effect that the curves of 
density profile extend further, but the overall behaviour of the systems 
does not change. Also, for the case $\sigma=1$ and $\sigma_{s}=1$ a third peak is formed
due to the layering effect as the density increases close to the rigid backbone.

\begin{figure*}                                                                                  
\begin{center}                                                                                   
\rotatebox{0}{\resizebox{!}{0.50\columnwidth}{%
  \includegraphics{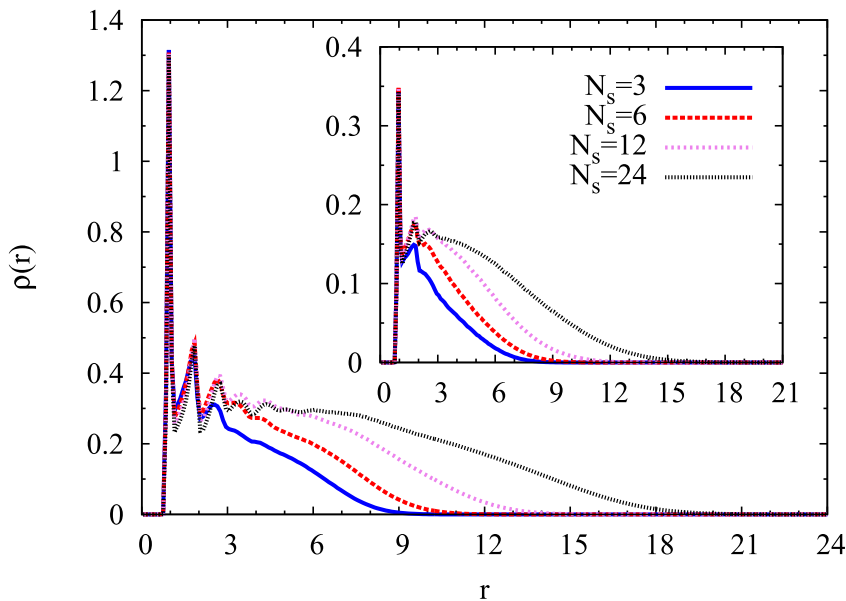}                                                                
}}                                                                                               
\end{center}                                                                                     
\caption{\label{fig4} 
(Colour online) Variation of the density profile $\rho(r)$ as a function 
of $r$ for the case of $\sigma=1.00$, $N_{b}=12$ and four different 
values of $N_{s}$ ($\sigma_{s}=0.5$). Inset
shows the corresponding results for $\sigma=0.25$.
}
\end{figure*}

Stretching of the flexible grafted bottle-brushes in the radial directions 
is more pronounced when one changes the length of the side chains of 
the grafted bottle-brushes. Figure~\ref{fig4} shows differences in the
density profile which result from the variation of $N_{s}$.
The increase of $N_{s}$ strengthens the extension of the density profile 
in the radial direction. The density now stays almost constant 
at rather high distances $r$ (the density remains constant even at longer
distances from the rigid backbone) and the resulting formation of the bottle-brush
brush resembles a homogeneous cylinder as it is that of figure ~\ref{fig2} (right part).
For $\sigma=1$ we see this  rather constant density up to a distance $7$ from the rigid 
backbone (disregarding of course the first two peaks, which are reminiscent of 
the layering effect observed for fluid particles in a box close to the 
wall). As $N_{s}$ increases, the density profile extends to 
higher distances from the rigid backbone.
This extension becomes a rather small effect as the 
grafting density $\sigma$ is low enough that the beads can not
fill the space close to the rigid backbone under good solvent conditions where
the chains stretch in the radial direction (inset). 

\begin{figure*}                                                                                                                
\begin{center}                                                                                                                 
\subfloat[][]{                                                                                                                 
\rotatebox{0}{\resizebox{!}{0.34\columnwidth}{%
  \includegraphics{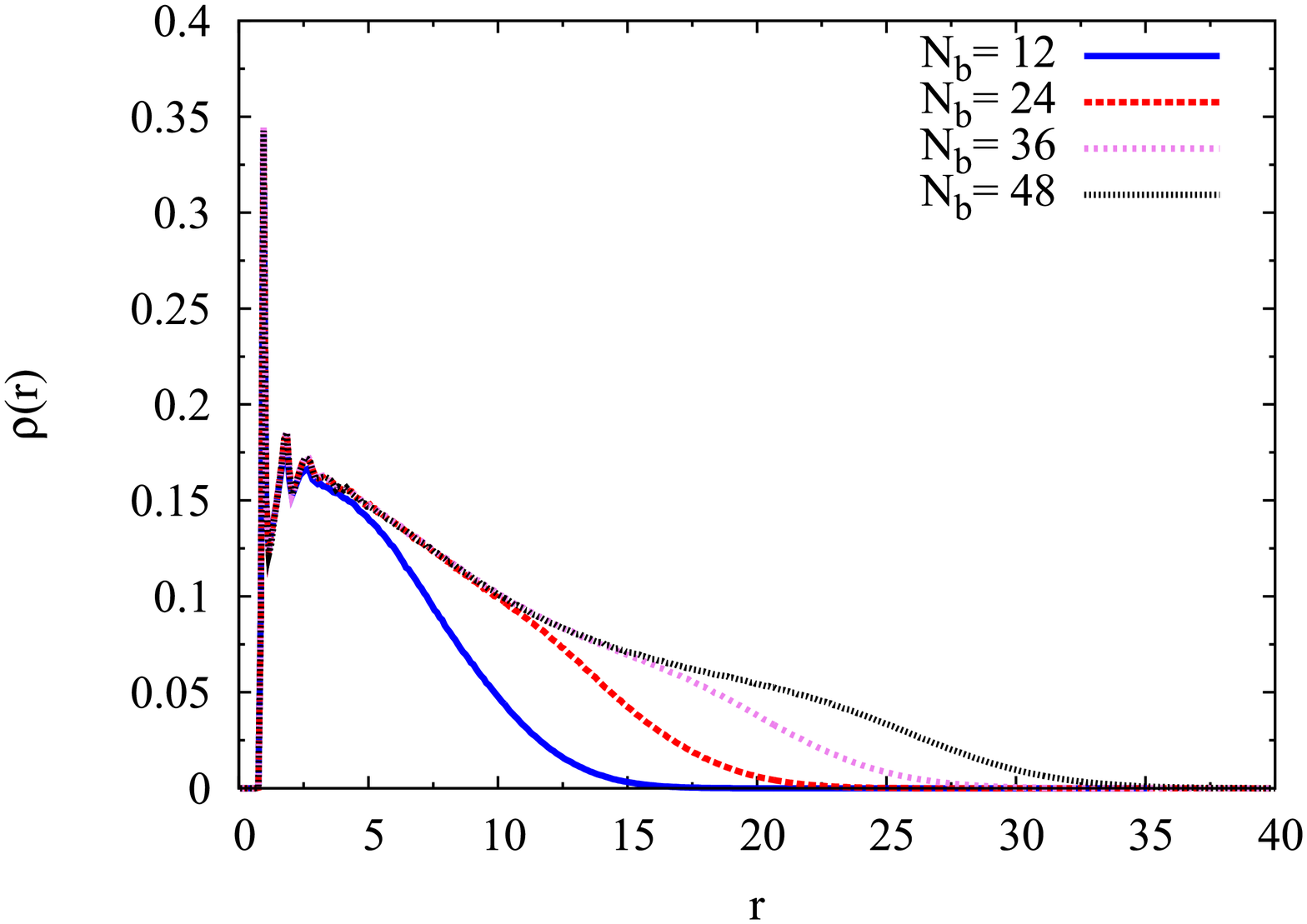}
}}}                                                                                                                            
\subfloat[][]{                                                                                                                 
\rotatebox{0}{\resizebox{!}{0.34\columnwidth}{%
  \includegraphics{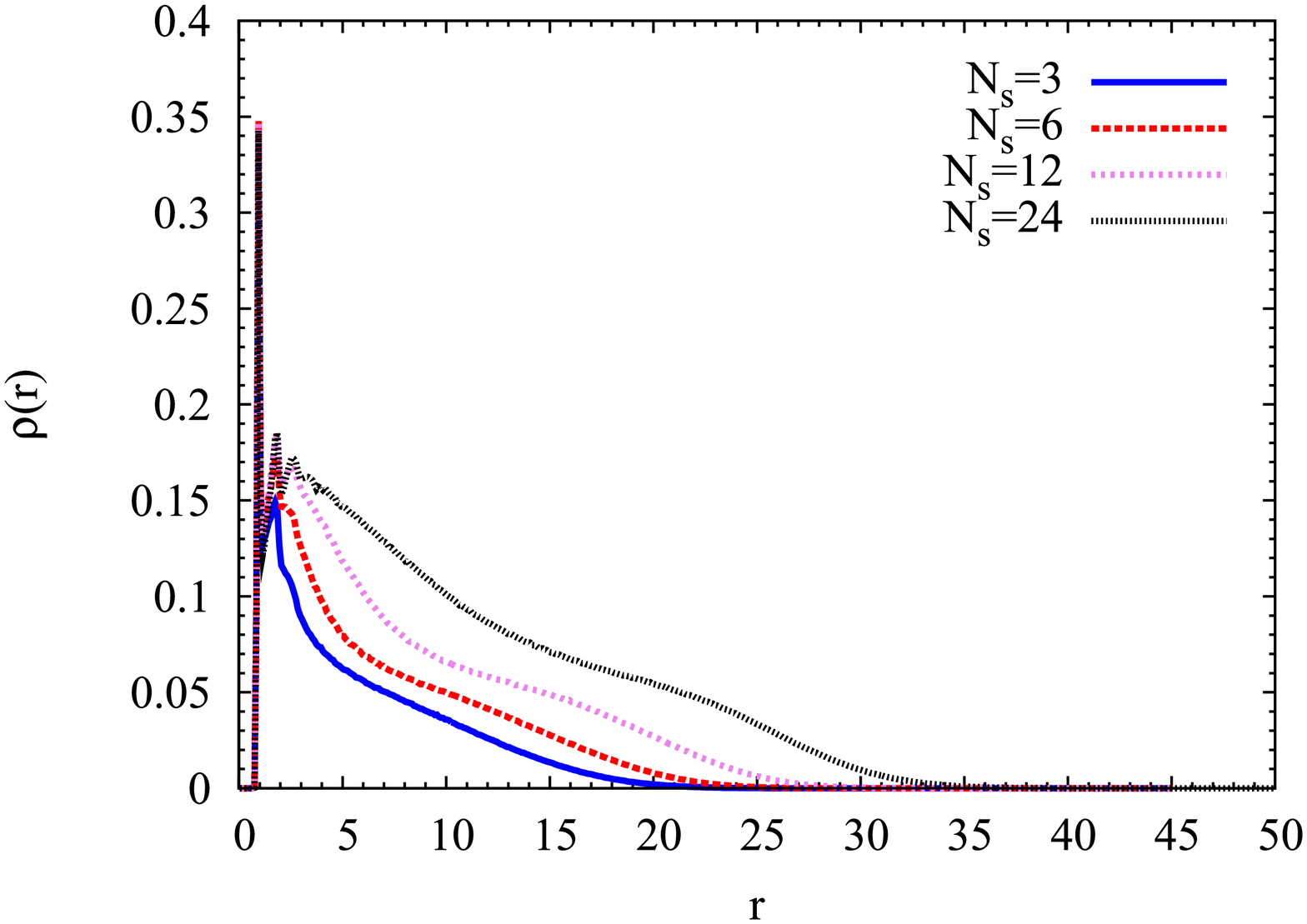}
}}}                                                                                                                            
\end{center}                                                                                                                   
\caption{\label{fig5}
(Colour online) Density profile $\rho(r)$ plotted versus the radial 
distance $r$. (a) The case $\sigma=0.25$, $N_{s}=24$ and four 
different brush chain lengths $N_{b}$ is shown. 
(b) Same as figure~\ref{fig4}, for the case $\sigma=0.25$, $N_{b}=48$ 
and different values of $N_{s}$. $\sigma_{s}=0.5$ in both cases (a) and (b).
}                                                                          
\end{figure*}

The variation of the density profile as a function of the radial distance $r$ 
with $N_{b}$ is shown in figure~\ref{fig5} (a). This typical 
graph is for the case of grafting density $\sigma=0.25$ and $N_{s}=24$. This
"low" grafting density allows us to discuss the particular effects avoiding
the structures where the density close to the rigid backbone is high.
At small distances $r$ from the rigid backbone, the difference in the 
density profiles can not be seen. However, at higher distances, the effect 
of long side chains ($N_{s}=24$) shows higher density at larger distances 
$r$. For $N_{b}<24$ the density profile looks as it is rather expected even
for a typical bottle-brush polymer.
But for $N_{b}>24$ (and $N_{s}=24$) the density profile shows a completely 
different behaviour when the length of the side chains ($N_{s}$) is also
rather high. This behavior is characteristic for the bottle-brush 
brush polymers. The density far from the rigid backbone is higher compared 
to a typical bottle-brush polymer. Without the side chains of the flexible 
brush, we would have expected and almost exponential decay in the 
density profile. However, as the length $N_{s}$ increases 
(figure~\ref{fig4}), the density in the outer region (far from the rigid 
backbone) of the bottle-brush brush polymer increases, and this 
difference is more pronounced, when the length $N_{b}$ is high enough 
in order to be far from the almost constant density region close to 
the rigid backbone (figure~\ref{fig5} (b)).

\begin{figure*}                                                                                  
\begin{center}                                                                                   
\rotatebox{0}{\resizebox{!}{0.50\columnwidth}{%
  \includegraphics{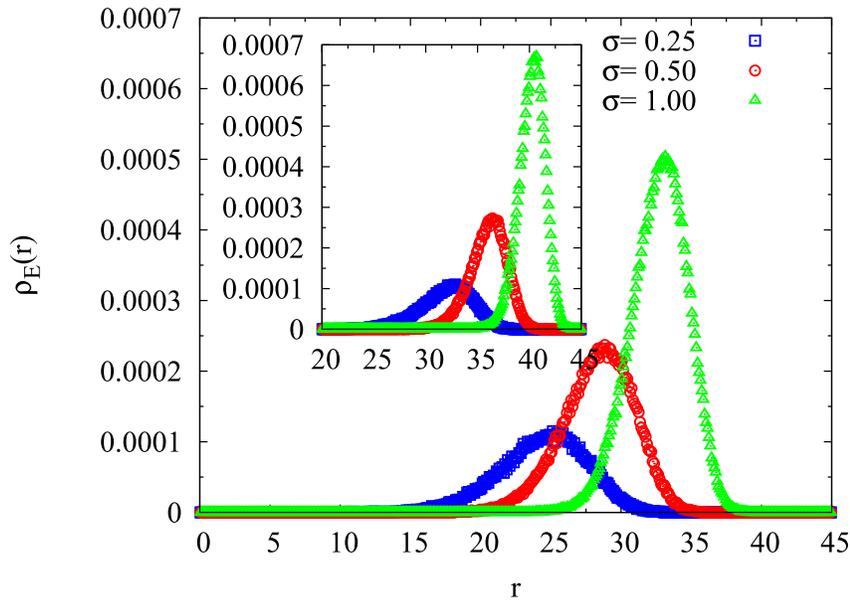}                                                                
}}                                                                                               
\end{center}                                                                                     
\caption{\label{fig6}                                                                            
(Colour online) Density profiles of the untethered end beads of the flexible 
backbones (bottle brush chains) $\rho_{E}(r)$ as a function of radial 
distance $r$ from the rigid backbone for different values of $\sigma$.
This typical illustration is for a system with $N_{b}=48$, 
$N_{s}=24$ and $\sigma_{s}=0.5$ ($\sigma_{s}=1.0$ in inset). 
}                                                                                                
\end{figure*}

More interesting is to study the behaviour of the ungrafted end beads 
of the bottle-brushes which are grafted on the rigid backbone in order to 
check how close they can come to the
rigid backbone. Figure~\ref{fig6} shows our results for the distribution of 
these end beads and the dependence with grafting density $\sigma$. 
Since $\sigma$ increases the peak of $\rho_{E}(r)$ becomes sharper
and thinner. This indicates that the end beads of the flexible backbone 
are moving within a narrower space in the radial direction and they rather 
never reach the region close to the rigid backbone. 
This effect clearly shows that the chains are stiffened due to the 
increase of the grafting density $\sigma$ and as a consequence, the 
peak is higher in the case of high $\sigma$. 
The stretching of the brush chains in the radial direction 
is also seen here and corroborates the results of figure~\ref{fig3}.
According to figure~\ref{fig6}, as $\sigma$ increases, the peaks are 
positioned at higher distances $r$. This effect is similar to the case 
of typical bottle-brush polymers where the increase of the grafting 
density induces stretching of the side chains in the radial directions 
and resulting in the increase of the zone which is not accessible to 
the end beads.                                                        
The increase of $\sigma_{s}$ strengthens the aforementioned effects, i.e.,                       
the brush chains become more stiffened and more stretched, although the
chains are under very good solvent conditions.                                       

\begin{figure*}                                                                                  
\begin{center}                                                                                   
\rotatebox{0}{\resizebox{!}{0.50\columnwidth}{%
  \includegraphics{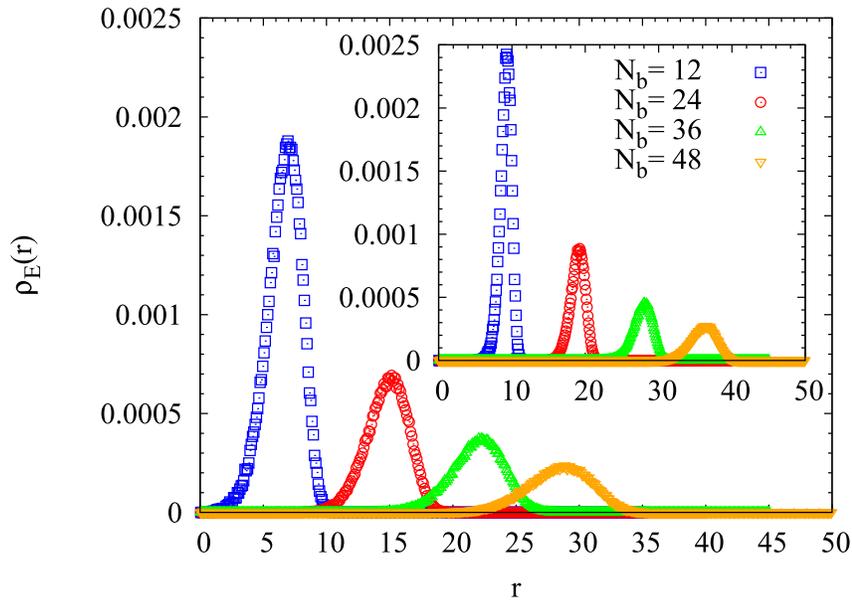}                                                                
}}                                                                                               
\end{center}                                                                                     
\caption{\label{fig7}                                                                            
(Colour online )Same as figure~\ref{fig6}, variation of density profile of the untethered end beads versus
radial distance $r$ for the case $\sigma=0.50$, $N_{s}=24$, and $\sigma_{s}=0.5$
($\sigma_{s}=1.0$ as inset), and different values of $N_{b}$.
}
\end{figure*}

The corresponding results for the end beads distribution for figure~\ref{fig5}, 
i.e., figure~\ref{fig7}, shows that the increase of $N_{b}$ shifts the 
peaks of the resulting density profiles to the right, while these 
peaks become less sharp, lower and smoother, showing that the flexible backbone becomes 
less mobile in the radial directions even for this case where the length 
of the side chains of the brush $N_{s}$ is rather high. 
Looking at figures ~\ref{fig6} and ~\ref{fig7}, we can see that all the changes
of parameters have a rather proportional effect on the properties. This points
out that carefull tuning of the structure of the bottle-brush brushes can result
from the design of our macromolecules, which can be performed in a variety of
ways, due to the variety of the parameters that describe their structure.

\begin{figure*}                                                                                                                
\begin{center}                                                                                                                 
\subfloat[][]{                                                                                                                 
\rotatebox{0}{\resizebox{!}{0.34\columnwidth}{%
  \includegraphics{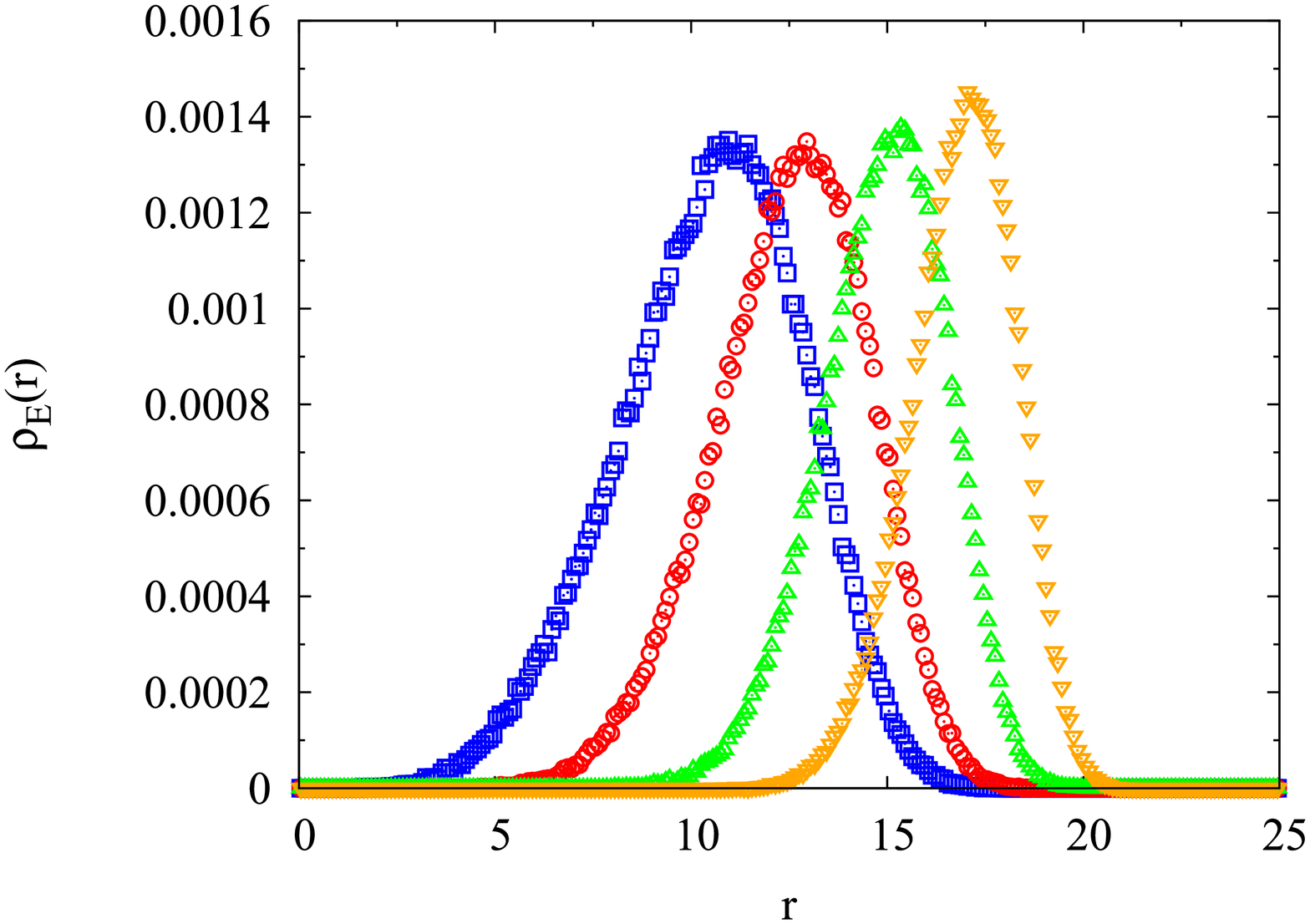}
}}}                                                                                                                            
\subfloat[][]{                                                                                                                 
\rotatebox{0}{\resizebox{!}{0.34\columnwidth}{%
  \includegraphics{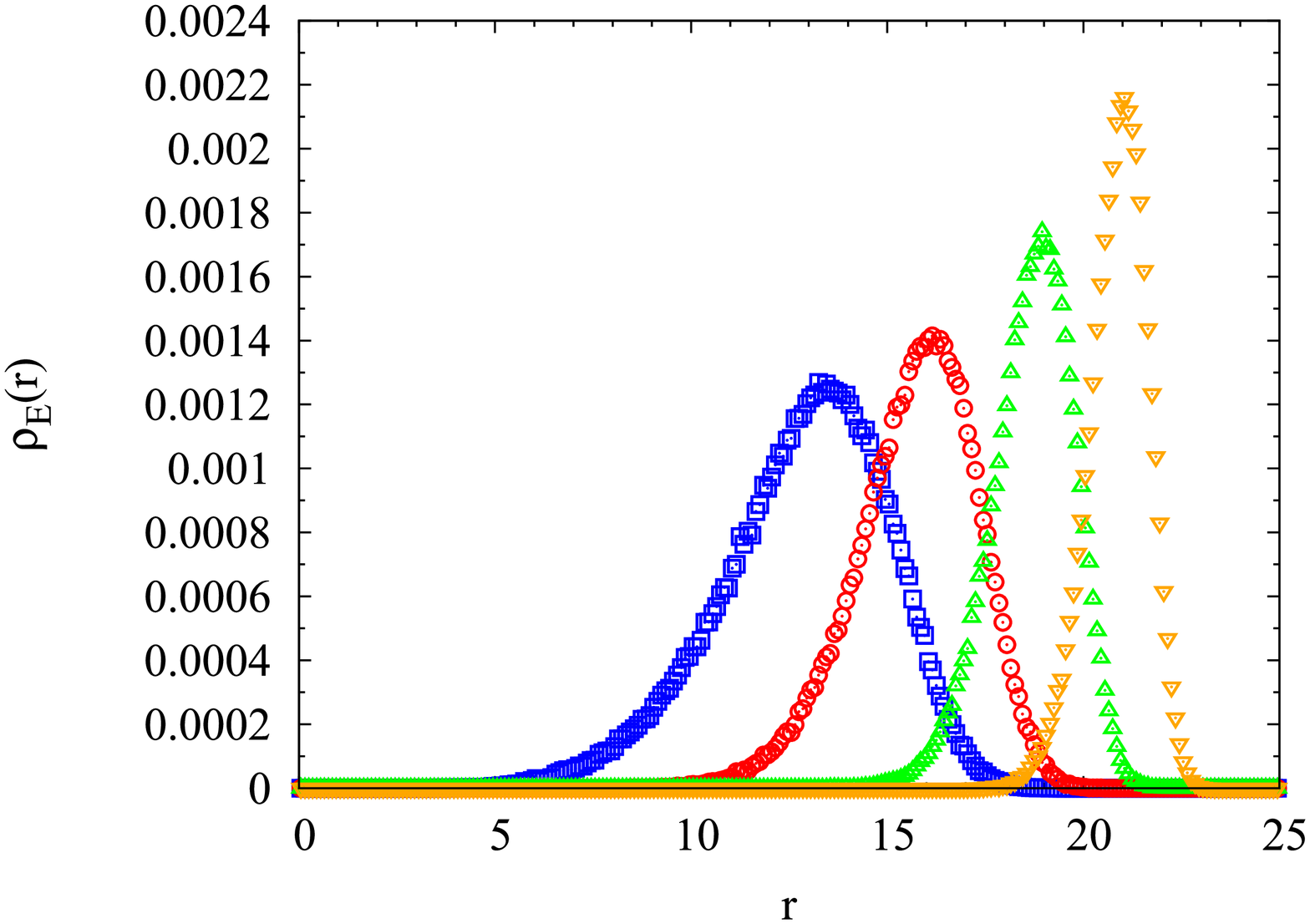} 
}}}                                                                                                                            
\end{center}                                                                                                                   
\caption{\label{fig8}
(Colour online) (a) Same as figure~\ref{fig6}, the flexible backbone 
(brush chains) ends $\rho_{E}(r)$ vs. $r$ for $\sigma=1.00$, $N_{b}=48$, 
and $\sigma_{s}=0.5$ ((b) $\sigma_{s}=1.0$), and different values of 
$N_{s}=3 (\Box),6 (\bigcirc), 12 (\bigtriangleup)$ and $24 (\bigtriangledown)$. 
}
\end{figure*}

When $\sigma_{s}$ increases, the differences in the curves are higher 
and one can see that the curves of the density profiles corresponding 
to different $N_{b}$ do not overlap any more. The latter effect points us out 
that the increase of the grafting density $\sigma_{s}$ has 
a significant effect and it is related analogously
to the probability for the flexible backbone untethered end of coming
close to the rigid backbone.
Figure~\ref{fig8} confirms this conclusion. When $\sigma_{s}=0.5$, 
for different values of $N_{s}$, we can almost observe the same 
height of the peaks in the density profiles for the end beads, the 
curves become slightly narrower, but the differences are not as 
pronounced as in the case of $\sigma_{s}=1.00$. 
This clearly shows that the change in the grafting density $\sigma_{s}$ 
has a significant effect on the bottle-brush brush polymers, 
which we could not distinguish in the overall radial density profiles.
We should also note that the increase of $\sigma_{s}$  introduces                                
noticeable differences in the height of the peaks, showing that this                             
parameter can play an important role in the resulting behaviour of
our system showing that the variations in $N_{s}$ result in more pronounced differences in the peaks
for high grafting density of side chains ($\sigma_{s}=1.0$), and they
now have a better correspondence to the values of $N_{s}$.

\begin{figure*}                                                                                  
\begin{center}                                                                                   
\rotatebox{0}{\resizebox{!}{0.50\columnwidth}{%
  \includegraphics{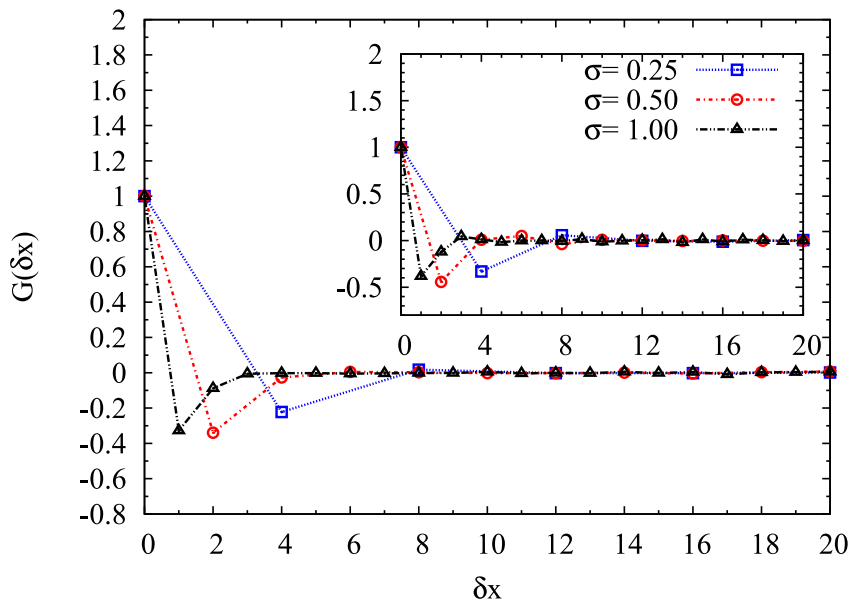}                                                                
}}                                                                                               
\end{center}                                                                                     
\caption{\label{fig9}                                                                            
(Colour online) The orientation correlation function $G(\delta x)$ as 
a function of the correlation distance $\delta x$ along hte rigid backbone
for a bottle-brush brush polymer with parameters
$N_{b}=12$, $N_{s}=3$ and $\sigma_{s}=0.5$ ($\sigma_{s}=1.0$ in inset)
for different grafting density $\sigma$. Only a range of $\delta x$ is
shown.
}
\end{figure*}

\begin{figure*}                                                                                                                
\begin{center}                                                                                                                 
\subfloat[][]{                                                                                                                 
\rotatebox{0}{\resizebox{!}{0.34\columnwidth}{%
  \includegraphics{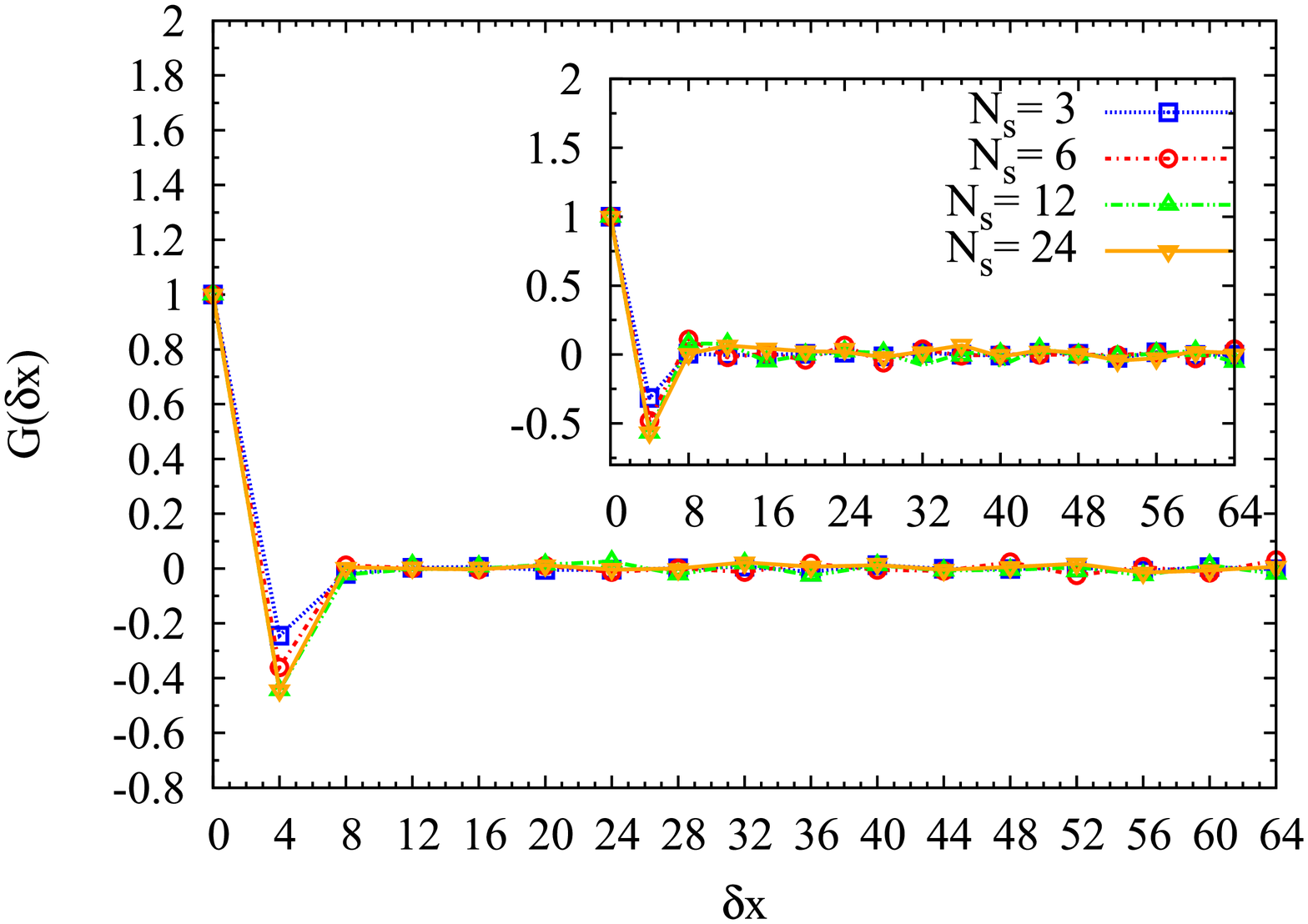}
}}}                                                                                                                            
\subfloat[][]{                                                                                                                 
\rotatebox{0}{\resizebox{!}{0.34\columnwidth}{%
  \includegraphics{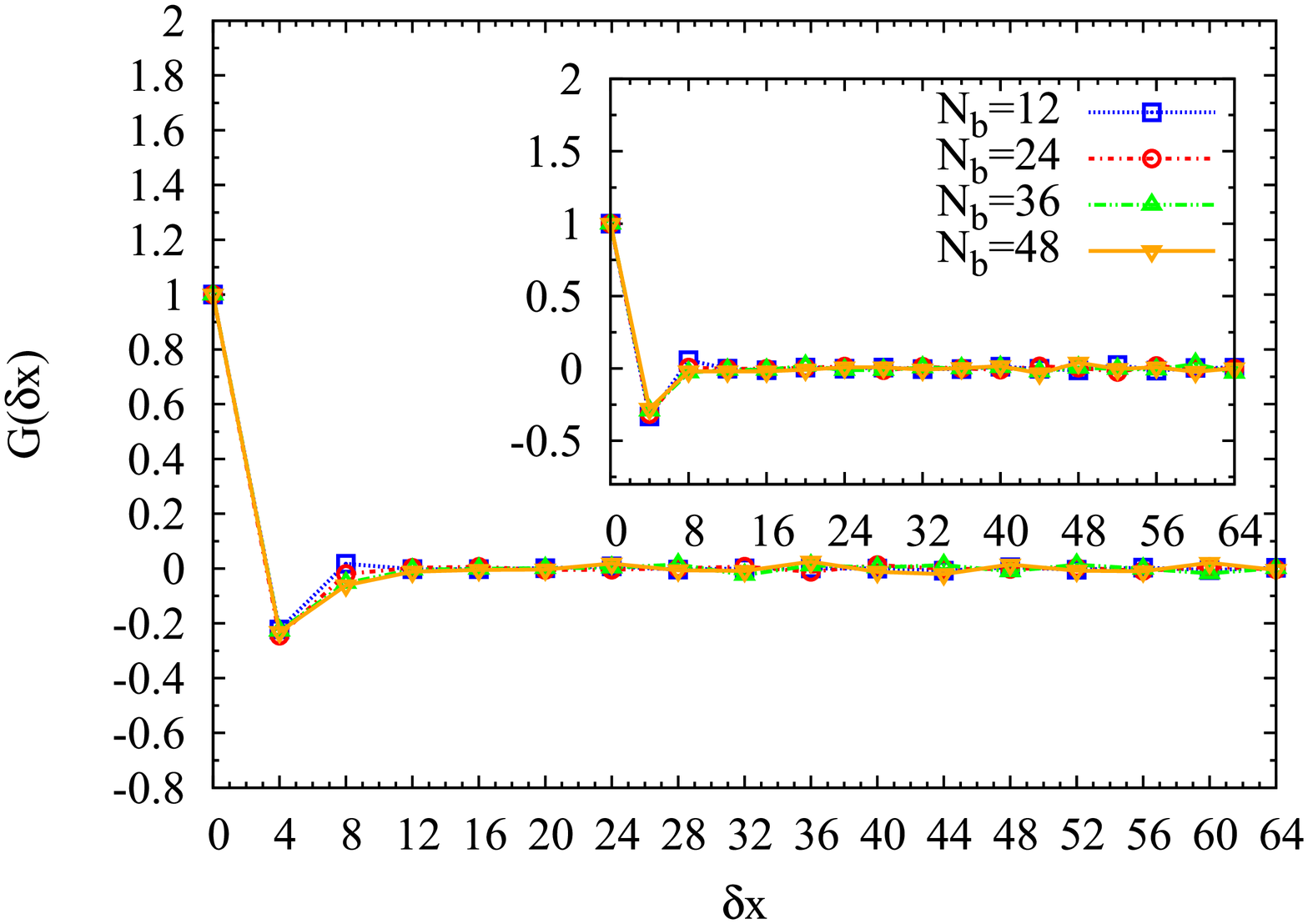}
}}}                                                                                                                            
\end{center}                                                                                                                   
\caption{\label{fig10}                                                                                                         
(Colour online)                                                                                                                
(a) The correlation function $G(\delta x)$ as a function of the distance
$\delta x$ along the rigid backbone
for a bottle-brush brush polymer with fixed grafted density $\sigma=0.25$,                                                     
$N_{b}=24$ and $\sigma_{s}=0.5$ for different side chain length $N_{s}$
(inset, $\sigma_{s}=1.0$).
(b) $G(\delta x)$ as a function of $\delta x$                                                 
for a bottle-brush brush polymer with fixed grafted density $\sigma=0.25$,                                                     
$N_{s}=3$ and $\sigma_{s}=0.5$ with different chain lengths $N_{b}$
(inset, $\sigma_{s}=1.0$). 
The first peak is at $1/\sigma$.
}                                                                                                                              
\end{figure*}

Figures~\ref{fig9} and \ref{fig10} show the correlation functions of 
orientations of different bottle-brushes in order to monitor 
the extent over which correlations between the different grafted 
brushes on the rigid backbone persist \cite{theodorakis_interplay_2010}. 
Such a property has been mainly considered for bottlebrush polymers with
two types of monomers ~\cite{theodorakis_interplay_2010} in order to 
characterize the extent over which a Janus structure exists. However,
in our case it would provide us the with information of the extent that these
compact objects (bottle-brushes grafted on the rigid backbone at high densities)
are affected by the presence of their neighboring bottle-brushes.
For each chain the vector pointing from the grafting site 
(the monomer on the rigid backbone where the brush is grafted) to the 
center of mass of the respective grafted bottle-brush is considered. Projecting this 
vector into the $yz$-plane and defining a unit vector $\vec{S_{i}}$ for the
$i$-th bottle-brush, we define the correlation function 
\begin{equation}\label{eqij}
G(\delta x)= \langle \vec{S_{i}} \cdot \vec{S_{j}} \rangle.                                     
\end{equation}
As we can see from figure~\ref{fig9}, the influence of the grafting density 
$\sigma$ for values $0.25$-$1.00$ is small, and only some difference 
can be observed when the value of $\sigma$ changes from $0.25$ to $0.50$. 
Further increase of the grafting density $\sigma$ seems to have a small
effect on the chains. These conclusions are also confirmed for other 
set of parameter $N_{b}$ and $N_{s}$ (not shown here). 
We should emphasize that these first peaks occur at a distance $1/\sigma$ 
as expected. 
Furthermore, the same behaviour is observed when the 
grafting density $\sigma_{s}$ is increased (inset in figure~\ref{fig9}). 
What we see is that the 
neighbour bottle-brushes influence mainly their first neighbours. Two 
neighbouring brushes grafted on the rigid backbone tend to orient in different 
directions due to the excluded volume interactions, whereas the second 
neighbours, the third, etc., are almost unaffected from the presence of the other 
grafted bottle-brushes on the rigid backbone. Only in the case that 
$\sigma=1.00$ some correlation is slightly detected also for the second 
neighbors. 
Deviations for $G(\delta x)$ from zero for third, forth, fifth, etc. neighboring
grafted brushes (figure ~\ref{fig9})
are within the statistical error. Of course, this is due to the
short chain lengths we have considered for our flexible brushes. When the
radius of gyration of the grafted brushes would be higher, these effects
play more important role.


For the set of parameters of figure~\ref{fig10}, we have varied parameters 
$N_{b}$ and $N_{s}$. Here we show only results for the case $\sigma=0.25$, in
order to discuss the particular effects. Similar conclusions are drawn for
the other cases.
Also the increase of $N_{s}$ (figure~\ref{fig10}(a)) shows that the correlations increase, 
but for $N_{s}>6$ the occurring differences are smaller. The variation 
of $N_{b}$ (figure~\ref{fig10} (b)) leads to the same conclusion. 
For the results of figures~\ref{fig9} and \ref{fig10}, we can further observe that the increase 
of $\sigma_{s}$ results in the increase of the correlations (which of 
course takes a negative value for neighboring brushes) along the rigid backbone, as the density 
of the monomers increases. The increase of $\sigma_{s}$ from $0.50$ 
to $1.00$ increases significantly the density of the monomers between 
the backbone of the grafted bottle-brushes, effect which is seen in 
the results of this correlation function. But even in the case of 
long lengths $N_{s}$ and $N_{b}$ and high grafting densities grafted 
bottle-brushes further apart form the first neighbors are hardly 
influenced. To conclude, for the the small lengths $N_{b}$ and $N_{s}$
considered here, the main effect comes from the variation of the grafting
density. For other combinations of parameters (not shown here in 
order to save space) the conclusions remain the
same.
%
%
\section{Summary}

In the present study, we performed molecular dynamic simulations 
of a standard bead-spring model of flexible bottle-brush polymers grafted onto 
a rigid backbone under good solvent conditions.                                                                      
Our investigation was based on the analysis of overall properties 
of interest,
i.e., the density profiles                                   
in the radial directions (which is the perpendicular direction to the rigid backbone                       
where the bottle-brush polymers are grafted with one of the ends of their                        
flexible backbone). We discussed that the increase of any of the parameters 
$\sigma$, $\sigma_{s}$, $N_{b}$, and $N_{s}$ results in the stretching of                        
the bottle-brushes in the radial directions although we are under
good solvent conditions, as has been also discussed for typical bottle-brushes
with a rigid or a flexible backbone.
We can also clearly see by analysis of the correlation function $G(\delta x)$ that mainly 
only first neighbor grafted bottle-brush chains are affected due                                 
to the excluded volume interactions between their monomers, even for our extreme 
cases where the grafted densities $\sigma$ and $\sigma_{s}$, 
and the lengths $N_{b}$
and $N_{s}$ are high. As the size of the grafted brushes increases, this effect is
expected to be more pronounced. The strength of the excluded volume effects between                   
two neighboring bottle-brush polymers grafted onto the rigid backbone                  
can be tuned in various ways, due to the high number of structural                             
parameters characterizing such macromolecules, compared to a typical bottle-brush 
polymer under good solvent conditions, where the only varying                              
parameter (considering also a rigid backbone) is the length of the grafted linear polymeric 
chains and the grafting density $\sigma$. We found that the variation of
these parameters affect in a proportional way the properties of the macromolecules.
That is a change of one parameter for all the grafted brushes has an analogous
measurable effect to the structural properties of these macromolecules, although
at a first glance their structure seems complicated to be studied by 
computer simulations.
Thus, the properties of these macromolecules can be tuned
efficiently by the change of parameters, of course when these are
in a good solvent. The lengths of the chains in 
our study were rather short, but we should stress here that such lengths could be
also accessible in the synthesis of bottle-brush brushes. We also find 
that for the range of parameters studied here, the untethered ends of the 
chains have not any possibility of being anywhere close to the rigid 
backbone even for small $\sigma$. The distance and the range that these
end beads (possible carriers of some particular substance) can access close
to the backbone can also efficiently be tuned, which is rather impossible to
control with such accuracy in a typical bottle-brush with flexible side chains.
The width of this zone (unaccessible to the end beads) 
depends strongly on the $N_{s}$ and $\sigma_{s}$ and on the increase 
of the density depending on the other parameters. 
We hope that our results will stimulate further study of these macromolecules 
and the work of analytical predictions.

\ackn{
Authors are grateful to Zentrum f\"{u}r Datenverarbeitung (ZDV) in uni-mainz 
for the computer facilities. P.E.T. is grateful for financial support 
by the Austrian Science Foundation within the SFB ViCoM (grant F41).}

%
%


\section*{References}

\begin{thebibliography}{99}

\bibitem{zhang_cylindrical_2005} Zhang M and M\"uller A H E 2005 
{\it J. Polym. Sci. Part A: Polym. Chem.} {\bf 43} 3461

\bibitem{koutalas_well-defined_2005} Koutalas G, Iatrou H, Lohse D and N. Hadjichristidis
2005 {\it Macromolecules} {\bf 38} 4996

\bibitem{subbotin_spatial_2007} Subbotin A V and Semenov A N 2007 
{\it Polymer Science, Ser. A} {\bf 49} 1328

\bibitem{ishizu_architecture_2008} Ishizu K , Murakami T and  TakanoT 2008 
{\it J. colloid and interface sci.} {\bf 322} 59

\bibitem{sheiko_cylindrical_2008} Sheiko S S , Sumerlin B S and Matyjaszewski K 2008
{\it Progr. Polym. Sci.} {\bf 33} 759

\bibitem{potemkin_comblike_2009} Potemkin I I and Palyulin V V 2009 
{\it Polymer Science, Ser. A} {\bf 51} 163

\bibitem{schappacher_newpolymerchain_2000} Schappacher M and Deffieux A 2000           
{\it Macromolecules} {\bf 33} 7371 

\bibitem{wintermantel_molecular_1996} Wintermatnel M, Gerle M, Fischer K,
Schmidt M, Wataoka I, Urakawa H, Kajiwara K and Tsukahara Y 1996 {\it Macromolecules}
{\bf 29}, 978


\bibitem{gunari_surfactant-induced_2008} Gunari N, Cong Y, Zhang B, Janshoff A 
and Schmidt M 2008 {\it Macromol. Rapid Commun.} {\bf 29} 821

\bibitem{matyjaszewski_atom_2001} Matyjaszewski K and Xia J 2001 
{\it Chem. Rev.} {\bf 101} 2921 

\bibitem{matyjaszewski_effect_2003} Matyjaszewski K, Qin S, Boyce J R, 
Shirvanyants D and Sheiko S S 2003 {\it Macromolecules} {\bf 36} 1843 


\bibitem{11} Rathgeber S, Pakula T, Matyjaszewski K and Beers K L 2005
{\it J. Chem. Phys.} {\bf 122} 124904                                                            
\bibitem{12} Zhang B, Gr\"ohn F, Pedersen J S, Fischer K,                                                            
Schmidt M 2006 {\it Macromolecules} {\bf 39} 8440                                                                                                                         
\bibitem{13} Rathgeber S, Pakula T, Wilk A, Matyajazewski K,
Lee H -I and Beers K L 2006 {\it Polymer} {\bf 47} 7318 

\bibitem{alexander_adsorption_1977} Alexander S 1977 
{\it J. Phys.(Paris)} {\bf 38} 983

\bibitem{de_gennes_conformations_1980} de Gennes P -G 1980 
{\it Macromolecules} {\bf 13} 1069

\bibitem{daoud_star_1982} Daoud M and Cotton J 1982 
{\it J. Phys.(Paris)} {\bf 43} 531

\bibitem{bug_theory_1987} Bug A L R,  Cates M E, Safran S A, and  WittenT A 1987
{\it J. Chem. Phys.} {\bf 87} 1824

\bibitem{milner_parabolic_1988} Milner S T,  Witten T A and  Cates M E 1988
{\it Europhys. Lett.} {\bf 5} 413

\bibitem{milner_effects_1989} Milner S T, Witten T A and Cates M E 1988
{\it Macromolecules} {\bf 21} 2610

\bibitem{milner_polymer_1991} Milner S T 1991                                               
{\it Science} {\bf 251} 905  

\bibitem{fredrickson_surfactant-induced_1993} Fredrickson G H 1993
{\it Macromolecules} {\bf 26} 2825

\bibitem{zhulina_scaling_1995} Zhulina E B and Vilgis T A 1995
{\it Macromolecules} {\bf 28} 1008

\bibitem{stephan_shape_2002} Stephan T, Muth S and Schmidt M 2002 
{\it Macromolecules} {\bf 35} 9857

\bibitem{li_new_2004} Li C, Gunari N, Fischer K, Janshoff A and Schmidt M 2004
{\it Angew. Chem. Int. Ed.} {\bf 43} 1101

\bibitem{theodorakis_microphase_2009} Theodorakis P E, Paul W, and Binder K 2009
{\it Europhys. Lett.} {\bf 88} 63002 

\bibitem{theodorakis_interplay_2010} Theodorakis P E, Paul W and Binder K 2010
{\it Macromolecules} {\bf 43} 5137.

\bibitem{theodorakis_pearl-necklace_2010} Theodorakis P E, Paul W and Binder K 2010 
{\it J. Chem. Phys.} {\bf 133} 104901

\bibitem{theodorakis_jcp_2011} Erukhimovich I, Theodorakis P E, Paul W and Binder K 2011 
{\it J. Chem. Phys.} {\bf 134} 054906

\bibitem{theodorakis_epje_2011} Theodorakis P E, Paul W and Binder K 2010 
{\it Eur. Phys. J. E} {\bf 34} 52

\bibitem{hsu_theodorakis_jcp_2011} Theodorakis P E, Hsu H-P, Paul W and Binder K 2011 
{\it J. Chem. Phys.} {\bf } 

\bibitem{hsu_characteristic_2010} Hsu H -P, Paul W, Rathgeber S and Binder K 2010
{\it Macromolecules} {\bf 43} 1592

\bibitem{hsu_one-_2007} Hsu H -P, Paul W and Binder K 2007
{\it Macromol. Theory Simul.} {\bf 16} 660

\bibitem{hsu_structure_2008} Hsu H -P, Paul W and Binder K 2008 
{\it J. Chem. Phys.} {\bf 129} 204904 

\bibitem{hsu_intramolecular_2006} Hsu H -P, Paul W and Binder K 2006 
{\it Europhys. Lett.} {\bf 76} 526

\bibitem{hsu_how_2009}  Hsu H -P, Binder K and Paul P 2009
{\it Phys. Rev. Lett.} {\bf 103} 198301

\bibitem{klein_chemistry:_2009} Klein J 2009
{\it Science} {\bf 323} 47

\bibitem{chang_structural_2009} Chang R, Kwak Y and Gebremichael Y J 2009
{\it Mol. Biol.} {\bf 391} 648

\bibitem{grest_molecular_1986} Grest G S and Kremer K 1986
{\it Phys. Rev. A} {\bf 33} 3628

\bibitem{graessley_excluded-volume_1999} Graessley W W, Hayward R C and Grest G S
{\it Macromolecules} {\bf 32} 3510

\bibitem{40} Grest G S and Murat M 1995  {\it in Monte Carlo and Molecular 
Dynamics Simulations in Polymer Science} 
Binder K Ed. p. 476. Oxford Univ. Press New York

\bibitem{grest_structure_1993} Grest G S and Murat M 1993 
{\it Macromolecules} {\bf 26} 3108

\bibitem{murat_polymers_1991} Murat M and Grest G S 1991 
{\it Macromolecules} {\bf 27} 704

\bibitem{86} Grest G S 1999 {\it Adv. Polym. Sci.} {\bf 138} 149 

\bibitem{murat_structure_1989} Murat M and Grest G S 1989
{\it Macromolecules} {\bf 22} 4054

\bibitem{grest_grafted_1994} Grest G S 1994 
{\it Macromolecules} {\bf 27} 418

\bibitem{lammps} http://lammps.sandia.gov/

\bibitem{plimpton_fast_1995} Plimpton S 1995
{\it J. Comput. Phys.} {\bf 117} 1



\end{thebibliography}
\end{document}